\documentclass[11pt]{article}
\usepackage{geometry}
\usepackage{amsmath, amssymb, amsthm}
\usepackage{graphicx}
\usepackage{bm}
\usepackage[linesnumbered,ruled]{algorithm2e}
\usepackage{fullpage}
\usepackage{mathtools}
\usepackage{url}
\usepackage{color} 

\newtheorem{lemma}{Lemma}

\newtheorem{claim}{Claim}

\title{A Simple Discrete Newton Method for the 
Quickest Transshipment Problem}
\author{Naoyuki Kamiyama\thanks{This work was 
supported by JSPS KAKENHI Grant Number JP24K14825.}
}
\date{Graduate School of Informatics\\ 
Kyoto University \\ 
Kyoto, Japan \\
\url{kamiyama@amp.i.kyoto-u.ac.jp}}

\begin{document}

\maketitle

\begin{abstract}
The quickest transshipment problem is a problem defined in  
a dynamic network with multiple sources and multiple sinks.
The goal of this problem is to send all the supplies 
from sources to sinks as quickly as possible. 
It is known that this problem can be solved in polynomial time. 
In particular, 
Schl{\"{o}}ter, Skutella, and Tran
proposed a strongly polynomial-time 
discrete Newton method for this problem. 
In this paper, we propose a simple polynomial-time  
discrete Newton method for the 
quickest transshipment problem. 
\end{abstract}

\section{Introduction} 

A dynamic network flow, which was introduced by Ford and
Fulkerson~\cite{FordF58,FordF62}, 
is a flow defined in
a dynamic network, i.e., a directed
graph with capacities and transit times on its arcs. 
The quickest transshipment problem is a problem defined in  
a dynamic network with multiple sources and multiple sinks.
The goal of this problem is to send all the supplies 
from sources to sinks as quickly as possible. 
It is known that the quickest transshipment problem
can be solved in polynomial time~\cite{HoppeT00,SchloterS17,SchloterST22}. 

Among the polynomial-time algorithms for the quickest transshipment problem, 
Schl{\"{o}}ter, Skutella, and Tran~\cite{SchloterST22}
proposed a strongly polynomial-time 
discrete Newton method for this problem. 
Compared to the single-sink case~\cite{Kamiyama19,Schloter18}, which is called the 
evacuation problem, 
the discrete Newton method for the case of multiple sinks and sources~\cite{SchloterST22}
is more complicated. 

In this paper, we propose a simple polynomial-time  
discrete Newton method for the 
quickest transshipment problem. 
Although the time complexity of the proposed algorithm is worse than that of 
\cite{SchloterST22}, 
the proposed algorithm has the following features.
\begin{itemize}
\item
In the algorithm of 
\cite{SchloterST22}, we have to compute the left derivative
of some function arising from the problem. 
On the other hand, in our algorithm, 
we do not need to compute the left derivative. 
\item
In each iteration of 
the algorithm of 
\cite{SchloterST22}, we check several 
candidates for the parameter value.
On the other hand, in the proposed algorithm, 
one candidate is sufficient. 
\end{itemize}

\paragraph{AI disclosure.}
The result in this paper was obtained 
by using GPT-6 Astra, and the author independently verified 
the mathematical correctness and 
reorganized the proof.
Furthermore, ChatGPT was used for proofreading.
The author 
takes full responsibility for the 
correctness and the content of this paper.

\section{Preliminaries}
\label{section:preliminaries} 

Let $\mathbb{R}$, $\mathbb{R}_+$, $\mathbb{Z}$,  
$\mathbb{Z}_+$, and 
$\mathbb{N}$ denote    
the sets of reals,  
non-negative reals, integers, non-negative integers, 
and positive integers, respectively.  
For each integer $z \in \mathbb{Z}_+$, 
we define 
$[z] \coloneqq \{1,2,\dots,z\}$ 
and 
$[z]_0 \coloneqq \{0,1,\dots,z\}$. 
Define $[0] \coloneqq \emptyset$. 
For each finite set $X$, each function $\eta \colon X \to \mathbb{R}$, and 
each subset $Y \subseteq X$, 
we define 
$\eta(Y) := \sum_{e \in Y}\eta(e)$.

The \emph{quickest transshipment problem} is defined as follows. 
In this problem, we are given a finite directed graph 
$D=(V,A)$ and non-empty disjoint subsets $S,T \subseteq V$.
Furthermore, 
we are given 
a capacity function $c \colon A \to \mathbb{N}$, 
a transit-time function $\tau \colon A \to \mathbb{Z}_+$, 
a supply function $b \colon S \cup T \to \mathbb{Z}$
such that 
$b(s) > 0$ for every vertex $s \in S$, 
$b(t) < 0$ for every vertex $t \in T$, and 
$b(S \cup T) = 0$. 
Define $n \coloneqq |V|$,  
$m \coloneqq |A|$, and 
$k \coloneqq |S \cup T|$. 
Assume that $D$ is weakly connected. 
Then we have $m \ge n - 1$. 

For each function $f \colon A \times \mathbb{Z}_+ \to \mathbb{R}_+$, 
each vertex $v \in V$, and each integer $\theta \in \mathbb{Z}_+$, 
we define 
\begin{equation*}
\partial f(v, \theta) := 
\sum_{a \in \delta_A(v)}\sum_{i = 0}^{\theta} f(a,i)
-
\sum_{a \in \varrho_A(v)}\sum_{i = 0}^{\theta - \tau(a)} f(a,i),
\end{equation*}
where $\delta_A(v)$ (resp.\ $\varrho_A(v)$) 
represents the set of 
arcs in $A$ leaving (resp.\ entering) $v$.
We assume that 
$\varrho_A(v) = \emptyset$ 
(resp.\ $\delta_A(v) = \emptyset$)
for every vertex $v \in S$ 
(resp. $v \in T$). 

For each integer $\Theta \in \mathbb{Z}_+$, 
a function $f \colon A \times \mathbb{Z}_+ \to \mathbb{R}_+$ 
is called a \emph{feasible dynamic flow} with respect to $\Theta$ 
if it satisfies the following conditions. 
\begin{description}
\setlength{\parskip}{0mm} 
\setlength{\itemsep}{1.2mm} 
\item[(D1)]
Let $a$ and $\theta$ be an arc in $A$ and an integer in $[\Theta]_0$, respectively. 
\begin{itemize}
\setlength{\parskip}{0mm} 
\setlength{\itemsep}{1.2mm} 
\item
If $\theta \le {\rm \Theta} - \tau(a)$, then 
$f(a, \theta) \le c(a)$.
\item
If $\theta > {\rm \Theta} - \tau(a)$, then 
$f(a, \theta) = 0$.
\end{itemize}
\item[(D2)]
For every vertex $v \in V \setminus (S \cup T)$ and 
every integer $\theta \in [\Theta]_0$, 
we have 
$\partial f(v,{\rm \theta}) \le 0$. 
\item[(D3)]
For every vertex $v \in S \cup T$, 
$\partial f(v,{\rm \Theta}) = b(v)$. 
\end{description}

In this paper, we assume that 
there exists an integer $\Theta \in \mathbb{Z}_+$ such that 
there exists a feasible dynamic flow with respect to 
$\Theta$. 
This can be checked in polynomial time as follows. 
First, we construct a bipartite graph whose vertex set 
is $S \cup T$. 
Each pair $(s,t) \in S \times T$ is connected by an edge if and only if 
$t$ is reachable from $s$ in $D$. 
Then the edges are directed from $S$ to $T$. 
Each vertex $v \in S \cup T$ has a supply or demand $b(v)$. 
Then we check whether there exists a flow from $S$ to $T$ 
satisfying all the supplies and demands. 

The goal of the quickest transshipment problem is 
to find the minimum integer $\Theta \in \mathbb{Z}_+$ 
such that there exists a feasible dynamic flow with respect to $\Theta$. 

\section{Useful Facts} 

In this section, we introduce 
facts that are needed in Section~\ref{section:algorithm}.

Assume that we are given a subset $X \subseteq S \cup T$ and 
an integer $\Theta \in \mathbb{Z}_+$.
Define ${\bf D}_{\Theta}(X)$ as 
the set of functions 
$f \colon A\times [\Theta]_0 \to \mathbb{R}_+$
satisfying (D1) and the following conditions (D4), (D5). 
\begin{description}
\setlength{\parskip}{0mm} 
\setlength{\itemsep}{1.2mm} 
\item[(D4)]  
Let $v$ and $\theta$ be a vertex in $V$ and an 
integer in $[\Theta]_0$, respectively. 
\begin{itemize}
\setlength{\parskip}{0mm} 
\setlength{\itemsep}{1.2mm} 
\item
If $v \in V \setminus X$, then 
$\partial f(v,\theta) \le 0$. 
\item
If $v \in X$, then 
$\partial f(v,\theta) \ge 0$. 
\end{itemize}
\item[(D5)]
For every vertex $v \in V \setminus (S \cup T)$, 
we have 
$\partial f(v, {\rm \Theta}) = 0$.
\end{description}
Recall that 
no arc in $A$ enters (resp., leaves) 
a vertex in $S$ (resp., $T$).
Thus, for every function $f \in {\bf D}_{\Theta}(X)$ and
every vertex $v \in (S \setminus X) \cup (T \cap X)$, 
$\partial f(v, {\rm \Theta}) = 0$ holds.  
Intuitively speaking, an element in ${\bf D}_{\Theta}(X)$ is a 
dynamic flow from $S \cap X$ to 
$T \setminus X$ without prescribed supplies. 
In addition, we define the function 
$o_{\Theta} \colon 2^{S \cup T} \to \mathbb{R}_+$ by   
\begin{equation*}
o_{\Theta}(X) := \max\Big\{\sum_{v \in X}\partial f(v,{\rm \Theta}) 
\ \Big| \ f \in {\bf D}_{\Theta}(X)\Big\}.
\end{equation*}
Intuitively speaking, $o_{\Theta}(X)$ is the maximum flow value 
of a dynamic flow from $S \cap X$ to 
$T \setminus X$. 
The following result was cited by Hoppe and Tardos~\cite{HoppeT00}
as personal communication.

\begin{lemma}[Klinz~{\cite[Theorem~5.1]{HoppeT00}}] \label{theorem:K00}
For each integer $\Theta \in \mathbb{Z}_+$, 
there exists a feasible dynamic flow 
with respect to $\Theta$ 
if and only if 
$o_{\Theta}(X) \ge b(X)$
for every subset $X \subseteq S \cup T$.
\end{lemma}

For each subset $X \subseteq S \cup T$, 
the goal of the \emph{quickest flow problem} $\mbox{\sc qf}(X)$ is 
to compute the minimum integer $\Theta \in \mathbb{Z}_+$ such that 
$o_{\Theta}(X) \ge b(X)$. 
It is known that this problem can be solved in 
$O(nm^2 \log^2 n)$ time~\cite{SahoS17}.\footnote{%
Precisely speaking, Saho and Shigeno~\cite{SahoS17} 
considered the continuous version of dynamic flows. 
However, it is not difficult to see that this algorithm is also valid for the discrete version.}

In what follows, let $v_S$, $v_T$ be distinct new vertices that are not contained in $V$. 
Then we define $U \coloneqq V \cup \{v_S,v_T\}$. 
For each vertex $s \in S$ (resp.\ $t \in T$), 
we define $a_{v_S,s}$ (resp.\ $a_{t,v_T}$) as an
arc from $v_S$ to $s$ (resp.\ $t$ to $v_T$). 
Furthermore, we define $a^{\ast}$ as the arc from $v_T$ to $v_S$. 

For each subset $X \subseteq S \cup T$, 
we introduce the following definitions. 
Define 
\begin{equation*}
E_X := A \cup \{a_{v_S,s} \mid s \in S \cap X\} \cup \{a_{t,v_T} \mid t \in T \setminus X\}
\cup \{a^{\ast}\}.
\end{equation*}
Define the directed graph $H_X$ as 
$(U,E_X)$.
For each subset $X \subseteq S \cup T$, 
a function $\xi \colon E_X \to \mathbb{R}_+$ is called a 
{\it feasible static flow} with respect to $X$ if 
it satisfies the following conditions (S1), (S2).
\begin{description}
\setlength{\parskip}{0mm} 
\setlength{\itemsep}{1.2mm} 
\item[(S1)]
For every arc $a \in A$, we have 
$\xi(a) \le c(a)$.
\item[(S2)]
For every vertex $v \in U$, 
we have 
$\xi(\delta_{E_X}(v)) = \xi(\varrho_{E_X}(v))$.
\end{description} 

For each subset $X \subseteq S \cup T$, 
we define ${\bf S}(X)$ as 
the set of feasible static flows with 
respect to $X$. 
For each subset $X \subseteq S \cup T$
and each real $\theta \in \mathbb{R}_+$, 
we define 
the cost vector $k_{\theta}^X \in \mathbb{R}^{E_X}$ as follows. 
For each arc $a \in A$,
we define $k_{\theta}^X(a):= -\tau(a)$. 
For each arc $a \in E_X \setminus (A \cup \{a^{\ast}\})$, 
we define 
$k_{\theta}^X(a):= 0$. 
Furthermore, 
we define 
$k_{\theta}^X(a^{\ast}):= {\rm \theta} + 1$. 
For each subset $X \subseteq S \cup T$
and each 
real $\theta \in \mathbb{R}_+$, we define the problem 
${\bf P}(X,\theta)$ as 
\begin{equation*}
\mbox{Maximize} \ \ 
\langle k_{\theta}^X, \xi \rangle \ \ 
\mbox{subject to} \ \ 
\xi \in {\bf S}(X). 
\end{equation*}
Since ${\bf S}(X)$ is non-empty and bounded 
(i.e., a non-empty polytope), 
there exists an extreme-point optimal solution to 
${\bf P}(X,\theta)$ (see, e.g., \cite[p.15]{GrotschelLS93}). 
Furthermore, for each subset $X \subseteq S \cup T$, 
we define $\rho_X \colon \mathbb{R}_+ \to \mathbb{R}$ by 
\begin{equation*}
\rho_X(\theta) \coloneqq 
\max\{\langle k_{\theta}^X, \xi \rangle
\mid \xi \in {\bf S}(X)\}. 
\end{equation*}
That is, $\rho_X(\theta)$ is the optimal objective value of 
${\bf P}(X,\theta)$.

\begin{lemma}[Ford and Fulkerson~\cite{FordF58,FordF62}] \label{theorem:FF58}
For every subset $X \subseteq S \cup T$ and 
every integer $\Theta \in \mathbb{Z}_+$, 
$o_{\Theta}(X)$ is equal to 
$\rho_X(\Theta)$.
\end{lemma}

It is known that, for every subset $X \subseteq S \cup T$, 
$\rho_X$
is piecewise linear, non-decreasing, and convex~\cite[p.93]{SchloterST22}. 
For each subset $X \subseteq S \cup T$ and 
each real $\theta \in \mathbb{R}_+ \setminus \{0\}$,
we define $\rho_X^-(\theta)$ 
as the left derivative of $\rho_X$ at $\theta$. 
It should be noted that, in this paper, 
$\rho_X^-(\theta)$ is not used in the proposed algorithm.

\section{Algorithm} 
\label{section:algorithm}

In this section, we give a new polynomial-time 
discrete Newton method for 
the quickest transshipment problem. 
For each integer $\theta \in \mathbb{Z}_+$,
we define the function $\Phi_{\theta} \colon 2^{S \cup T} \to \mathbb{R}$
by 
\begin{equation*}
\Phi_{\theta}(X) \coloneqq o_{\theta}(X) - b(X).
\end{equation*}
Notice that, for every integer $\theta \in \mathbb{Z}_+$, 
$\Phi_{\theta}(\emptyset) = 0$.
Define the function $d \colon \mathbb{Z}_+ \to \mathbb{R}$ by 
\begin{equation*}
d(\theta) \coloneqq \min_{X \subseteq S \cup T}
\Phi_\theta(X).
\end{equation*}
Then Lemma~\ref{theorem:K00}
implies that,
for each integer $\Theta \in \mathbb{Z}_+$, 
there exists a feasible dynamic flow 
with respect to $\Theta$ 
if and only if 
$d(\Theta) = 0$. 
Furthermore, we define $p \coloneqq |S|\cdot |T|$. 
Then $p \le k^2$. 

The proposed method is given in Algorithm~\ref{alg:main}. 

\begin{algorithm}[ht]
Set $\lambda \coloneqq 1$. Define $\theta_1 \coloneqq 0$.\\
\While{$d(\theta_{\lambda}) < 0$}
{
  Find a minimizer $X_{\lambda}$ of $\Phi_{\theta_{\lambda}}$. \\
  Let $r_{\lambda}$ be the optimal solution of $\mbox{\sc qf}(X_{\lambda})$. \\
  Define $q_{\lambda} \coloneqq r_{\lambda} + p(r_{\lambda} - \theta_{\lambda})$. \\
  If $d(q_{\lambda}) < 0$, then we define $\theta_{{\lambda}+1} \coloneqq q_{\lambda}$.
  Otherwise, we define $\theta_{\lambda + 1} \coloneqq r_{\lambda}$.\\
  Update ${\lambda} \coloneqq {\lambda}+1$. 
}
Output $\theta_{\lambda}$, and halt. 
\caption{Discrete Newton method for the quickest transshipment problem}
\label{alg:main}
\end{algorithm}

In the course of Algorithm~\ref{alg:main}, 
if $d(\theta_{\lambda}) < 0$, then 
$\Phi_{\theta_{\lambda}}(X_{\lambda}) < 0$ and 
$\Phi_{\theta_{\lambda + 1}}(X_{\lambda}) \ge 0$.
This and 
Lemma~\ref{theorem:K00}
imply that the number of iterations in 
Algorithm~\ref{alg:main} is bounded by 
$2^k$. 

Define $\Theta^{\ast}$ as the optimal solution to 
the quickest transshipment problem. 
Then Lemma~\ref{theorem:K00}
implies that, in the course of Algorithm~\ref{alg:main}, 
we have $\Theta^{\ast} \ge r_{\lambda}$. 
Furthermore, we define $\theta_{\lambda + 1} \coloneqq q_{\lambda}$ only 
when $d(q_{\lambda}) < 0$, i.e., 
$q_{\lambda} < \Theta^{\ast}$.  
Thus, the output of 
Algorithm~\ref{alg:main} is $\Theta^{\ast}$.

\subsection{Time complexity} 

In this section, we prove that 
Algorithm~\ref{alg:main} is a polynomial-time algorithm. 
Since we can evaluate $o_{\theta_{\lambda}}(X)$ for each subset $X \subseteq S\cup T$ 
in polynomial time (e.g., \cite{Orlin93}) and 
$o_{\theta_{\lambda}}(\cdot)$ is submodular~\cite{HoppeT00}, 
we can find a minimizer of $\Phi_{\theta_{\lambda}}$ 
in polynomial time by using any polynomial-time 
submodular function minimization algorithm (e.g., \cite{IwataFF01,LeeSW15,Schrijver00}). 
Furthermore, Step~4 can be done in  
polynomial time~\cite{SahoS17}. 
Thus, we consider the number of iterations in 
Algorithm~\ref{alg:main}. 

Assume that Algorithm~\ref{alg:main} halts when $\lambda = \Pi + 1$ and 
$\Pi \ge 1$.
For every integer $\ell \in [\Pi]$, 
since $\Phi_{\theta_{\ell}}(X_{\ell}) < 0$ and 
$\Phi_{r_{\ell}}(X_{\ell}) \ge 0$,
we have $r_{\ell} > \theta_{\ell}$. 
Define 
\begin{equation*}
I_{\rm A} \coloneqq \{\ell \in [\Pi] \mid \theta_{\ell+1} = q_{\ell}\}, \ \ 
I_{\rm R} \coloneqq [\Pi] \setminus I_{\rm A}.
\end{equation*} 
Notice that since 
$d(\theta_{\Pi+1}) \ge 0$, we have $\Pi \in I_{\rm R}$. 

First, we evaluate the size of $I_{\rm A}$. 
For each subset $X \subseteq S \cup T$, we define
\begin{equation*}
{\rm Pair}(X) \coloneqq 
\{(s,t) \in S \times T \mid s \in X, t \notin X\}. 
\end{equation*}

\begin{lemma} \label{lemma:non-empty_pair}
For every integer $\ell \in [\Pi]$, 
${\rm Pair}(X_{\ell}) \neq \emptyset$. 
\end{lemma}
\begin{proof}
Let $\ell$ be an integer in $[\Pi]$. 
Then $d(\theta_{\ell}) = o_{\theta_{\ell}}(X_{\ell}) - b(X_{\ell}) < 0$. 
Thus, since $o_{\theta_{\ell}}(X_{\ell}) \ge 0$, 
$b(X_{\ell}) > 0$.
This implies that 
$S \cap X_{\ell} \neq \emptyset$. 
Furthermore, if $T \subseteq X_{\ell}$, then 
$o_{r_{\ell}}(X_{\ell}) = 0$. 
However, this contradicts the fact that 
$o_{r_{\ell}}(X_{\ell}) \ge b(X_{\ell}) > 0$. 
Thus, ${\rm Pair}(X_{\ell}) \neq \emptyset$. 
\end{proof} 

\begin{lemma}[{Schl\"oter, Skutella, and Tran~\cite[Equation~(4.2)]{SchloterST22}}] \label{lemma:SchloterST22_4.2}
Let $X$, $\theta$ be a subset of $S \cup T$ and 
a real in $\mathbb{R}_+ \setminus \{0\}$, respectively. 
In addition, we assume that 
${\rm Pair}(X) \neq \emptyset$. 
Then there exists an element $(s,t) \in {\rm Pair}(X)$
such that 
\begin{equation*}
\forall Y \subseteq S \cup T \ \ \mbox{\rm s.t.} \ 
(s,t) \in {\rm Pair}(Y) \colon 
\rho^-_Y(\theta) \ge 
\frac{\rho^-_X(\theta)}{p}. 
\end{equation*}
\end{lemma}

\begin{lemma} \label{lemma:accept}
$|I_{\rm A}| \le p$. 
\end{lemma}
\begin{proof}
Lemma~\ref{lemma:non-empty_pair} implies that, 
for each element $\ell \in I_{\rm A}$, 
${\rm Pair}(X_{\ell}) \neq \emptyset$. 
Thus, 
Lemma~\ref{lemma:SchloterST22_4.2} implies that, 
for each element $\ell \in I_{\rm A}$, 
there exists an element $(s_{\ell},t_{\ell}) \in {\rm Pair}(X_{\ell})$ 
such that 
\begin{equation*}
\forall Y \subseteq S \cup T \ \ \mbox{\rm s.t.} \ 
(s_{\ell},t_{\ell}) \in {\rm Pair}(Y) \colon 
\rho^-_Y(r_{\ell}) \ge 
\frac{\rho^-_{X_{\ell}}(r_{\ell})}{p}. 
\end{equation*}

\begin{claim} \label{claim_1:lemma:accept} 
$(s_{\ell},t_{\ell}) \notin {\rm Pair}(X_{\zeta})$
holds 
for every pair of elements $\ell,\zeta \in I_{\rm A}$
such that $\ell < \zeta$.
\end{claim}

If we can prove Claim~\ref{claim_1:lemma:accept}, 
then this lemma can be proved as follows. 
Since $(s_{\ell},t_{\ell}) \in {\rm Pair}(X_{\ell})$ for every element $\ell \in I_{\rm A}$, 
Claim~\ref{claim_1:lemma:accept} implies that 
$(s_{\ell},t_{\ell}) \neq (s_{\zeta},t_{\zeta})$ for every pair of distinct 
elements $\ell,\zeta \in I_{\rm A}$. 
Assume that $|I_{\rm A}| > p$.
Let $\zeta$ be the maximum integer in $I_{\rm A}$. 
Since 
$|S \times T| \le p$,   
\begin{equation*}
\{(s_{\ell},t_{\ell}) \mid \ell \in I_{\rm A} \setminus \{\zeta\}\} = S \times T. 
\end{equation*}
Thus, since it follows from Claim~\ref{claim_1:lemma:accept}
that
$(s_{\ell},t_{\ell}) \notin {\rm Pair}(X_{\zeta})$
for every element $\ell \in I_{\rm A} \setminus \{\zeta\}$, 
we 
have ${\rm Pair}(X_{\zeta}) = \emptyset$.
However, this contradicts Lemma~\ref{lemma:non-empty_pair}. 

\begin{proof}[Proof of Claim~\ref{claim_1:lemma:accept}]
Let $\ell,\zeta$ be elements in $I_{\rm A}$ such that 
$\ell < \zeta$. 

Since $r_{\ell} > \theta_{\ell}$ and $\rho_{X_{\ell}}$ is convex, 
the definition of the left derivative implies that 
\begin{equation} \label{eq_1:claim_1:lemma:accept}
\rho_{X_{\ell}}^-(r_{\ell}) \ge 
\frac{\rho_{X_{\ell}}(r_{\ell}) - \rho_{X_{\ell}}(\theta_{\ell})}{r_{\ell} - \theta_{\ell}}. 
\end{equation}
Recall that the definition of $r_{\ell}$ implies that 
$\rho_{X_{\ell}}(r_{\ell}) = o_{r_{\ell}}(X_{\ell}) \ge b(X_{\ell})$.
Thus, 
\eqref{eq_1:claim_1:lemma:accept}
implies that 
\begin{equation*}
\rho_{X_{\ell}}^-(r_{\ell}) \ge 
\frac{b(X_{\ell}) - \rho_{X_{\ell}}(\theta_{\ell})}{r_{\ell} - \theta_{\ell}}
= 
\frac{b(X_{\ell}) - o_{\theta_{\ell}}(X_{\ell})}{r_{\ell} - \theta_{\ell}}
= 
\frac{-\Phi_{\theta_{\ell}}(X_{\ell})}{r_{\ell} - \theta_{\ell}}
= 
\frac{-d(\theta_{\ell})}{r_{\ell} - \theta_{\ell}}.
\end{equation*}
where 
the third equality follows from the fact that 
$X_{\ell}$ is a minimizer of $\Phi_{\theta_{\ell}}$.
Thus,    
\begin{equation} \label{eq_2:claim_1:lemma:accept}
\rho_{X_{\ell}}^-(r_{\ell}) \cdot (r_{\ell} - \theta_{\ell}) \ge - d(\theta_{\ell}). 
\end{equation}

Assume that $(s_{\ell},t_{\ell}) \in {\rm Pair}(X_{\zeta})$.
Since $\rho_{X_{\zeta}}$ is non-decreasing, 
\begin{equation} \label{eq_3:claim_1:lemma:accept}
\Phi_{r_{\ell}}(X_{\zeta}) 
= \rho_{X_{\zeta}}(r_{\ell}) - b(X_{\zeta})
\ge \rho_{X_{\zeta}}(\theta_{\ell}) - b(X_{\zeta})
= 
\Phi_{\theta_{\ell}}(X_{\zeta}) 
\ge d(\theta_{\ell}). 
\end{equation}
The convexity of $\rho_{X_{\zeta}}$ implies that 
\begin{equation*}
\Phi_{q_{\ell}}(X_{\zeta}) - \Phi_{r_{\ell}}(X_{\zeta})
= o_{q_{\ell}}(X_{\zeta}) - o_{r_{\ell}}(X_{\zeta}) 
= \rho_{X_{\zeta}}(q_{\ell}) - \rho_{X_{\zeta}}(r_{\ell}) 
\ge 
\rho_{X_{\zeta}}^-(r_{\ell}) \cdot (q_{\ell} - r_{\ell}). 
\end{equation*}
Thus, since $(s_{\ell},t_{\ell}) \in {\rm Pair}(X_{\zeta})$ and 
$q_{\ell} - r_{\ell} = p(r_{\ell} - \theta_{\ell})$, 
Lemma~\ref{lemma:SchloterST22_4.2} implies that 
\begin{equation} \label{eq_4:claim_1:lemma:accept}
\begin{split}
\Phi_{q_{\ell}}(X_{\zeta}) & \ge 
\Phi_{r_{\ell}}(X_{\zeta}) + \rho_{X_{\zeta}}^-(r_{\ell}) \cdot (q_{\ell} - r_{\ell}) 
 \ge 
\Phi_{r_{\ell}}(X_{\zeta}) + \frac{\rho_{X_{\ell}}^-(r_{\ell})}{p} \cdot (q_{\ell} - r_{\ell}) \\
& \ge 
\Phi_{r_{\ell}}(X_{\zeta}) + \rho_{X_{\ell}}^-(r_{\ell}) \cdot (r_{\ell} - \theta_{\ell}) 
 \ge 
d(\theta_{\ell}) - d(\theta_{\ell}) = 0, 
\end{split} 
\end{equation}
where the last inequality follows from 
\eqref{eq_2:claim_1:lemma:accept} and 
\eqref{eq_3:claim_1:lemma:accept}. 

Since $\rho_{X_{\zeta}}$ is non-decreasing and 
$\theta_{\zeta} \ge \theta_{\ell+1} = q_{\ell}$, 
$o_{\theta_{\zeta}}(X_{\zeta}) \ge o_{q_{\ell}}(X_{\zeta})$. 
Thus, 
since $\Phi_{\theta_{\zeta}}(X_{\zeta}) \ge \Phi_{q_{\ell}}(X_{\zeta})$, 
\eqref{eq_4:claim_1:lemma:accept} implies that 
$\Phi_{\theta_{\zeta}}(X_{\zeta}) \ge 0$.
However, since $d(\theta_{\zeta}) = \Phi_{\theta_{\zeta}}(X_{\zeta})$, 
this contradicts the fact that $\zeta \in [\Pi]$.  
This completes the proof. 
\end{proof} 
As explained above, 
Claim~\ref{claim_1:lemma:accept} completes the proof of 
this lemma. 
\end{proof}

Next, we evaluate the size of $I_{\rm R}$. 
For each integer $z \in \mathbb{N}$, we define $\kappa(z) \coloneqq 4z \log_2 z + 3$. 

\begin{lemma}[{Radzik~\cite[Lemma~3]{Radzik13}}] \label{lemma:Radzik13}
Let $\tau$ be a positive integer. 
Assume that we are given a vector $w \in \mathbb{R}^{\tau}_+$
and vectors $z_1,z_2,\dots,z_{\delta} \in \{-1,0,1\}^{\tau}$ such that 
$\langle z_i,w\rangle >0$ for every integer $i \in [\delta]$ and 
$\langle z_i,w\rangle \ge 2 \cdot \langle z_{i+1},w\rangle$ for every integer 
$i \in [\delta-1]$.
Then we have $\delta \le \kappa(\tau)$. 
\end{lemma}

\begin{lemma} \label{lemma:interval_cover} 
Let $\tau$ be a positive integer. 
Assume that we are given a vector $w \in \mathbb{R}^{\tau}_+$.
Define 
\begin{equation*}
\mathcal{A} \coloneqq \{\langle z,w \rangle 
\mid \langle z,w \rangle > 0, \ z \in \{-1,0,1\}^{\tau} \}. 
\end{equation*}
Assume that $\mathcal{A} \neq \emptyset$. 
Then there exist positive reals $h_1,h_2,\dots,h_{\delta}$ such that 
$\delta \le \kappa(\tau)$ and 
\begin{equation*}
\mathcal{A} \subseteq 
\bigcup_{i=1}^{\delta}\{x \in \mathbb{R}_+ \mid h_i/2 < x \le h_i\}.
\end{equation*} 
\end{lemma}
\begin{proof}
We recursively define $h_i$ as follows. 
Define $h_1$ as the maximum element in $\mathcal{A}$. 
For each integer $i \ge 2$, if 
there exists an element $x \in \mathcal{A}$ such that 
$x \le h_{i-1}/2$, then 
we define 
\begin{equation*}
h_i \coloneqq \max \{x \in \mathcal{A} \mid x \le h_{i-1}/2\}.
\end{equation*}
Since $\mathcal{A}$ is finite, this procedure terminates. 
Furthermore, Lemma~\ref{lemma:Radzik13} 
implies that the number of iterations is at most $\kappa(\tau)$. 
This completes the proof. 
\end{proof} 

\begin{lemma}[{See, e.g.,~\cite[pp.~444--447]{AhujaMO93}}] \label{lemma:AhujaMO93}
Let $X$ be a subset of $S \cup T$, and 
let $\xi$ be an extreme point in ${\bf S}(X)$.
Then there exists a vector
$\varepsilon_{\xi} \in \{-1,0,1\}^{E_X \times A}$ such that 
\begin{equation*}
\forall g \in E_X \colon 
\xi(g) = \sum_{a \in A}\varepsilon_{\xi}(g,a) \cdot c(a).
\end{equation*}
\end{lemma} 

\begin{lemma} \label{lemma:geometric}
For every element $\ell \in I_{\rm R}$, 
\begin{equation*}
\Theta^{\ast} - r_{\ell} \le \frac{p}{p+1}(\Theta^{\ast} - \theta_{\ell}). 
\end{equation*}
\end{lemma} 
\begin{proof}
Let $\ell$ be an element in $I_{\rm R}$.
Then 
since $d(q_{\ell}) \ge 0$, $\Theta^{\ast} \le q_{\ell}$. 
Thus, 
\begin{equation*}
\Theta^{\ast} \le q_{\ell} 
= r_{\ell} + p(r_{\ell} - \theta_{\ell})
= \theta_{\ell} + (p+1)(r_{\ell} - \theta_{\ell}).
\end{equation*}
This implies that 
\begin{equation*}
\Theta^{\ast} - r_{\ell} 
= (\Theta^{\ast} - \theta_{\ell}) - (r_{\ell} - \theta_{\ell})
\le (\Theta^{\ast} - \theta_{\ell}) - \frac{(\Theta^{\ast} - \theta_{\ell})}{p+1}
= \frac{p}{p+1} (\Theta^{\ast} - \theta_{\ell}).
\end{equation*}
This completes the proof. 
\end{proof} 

For each element $\ell \in I_{\rm R}$, we define 
$\xi_{\ell}$ as an extreme-point optimal solution  
to ${\bf P}(X_{\ell},r_{\ell})$. It 
follows from 
Lemma~\ref{lemma:AhujaMO93}
that there exists a vector
$\varepsilon_{\xi_{\ell}} \in \{-1,0,1\}^{E_{X_{\ell}} \times A}$ such that 
\begin{equation*}
\forall g \in E_{X_{\ell}} \colon 
\xi_{\ell}(g) = \sum_{a \in A}\varepsilon_{\xi_{\ell}}(g,a) \cdot c(a).
\end{equation*}
For each element $\ell \in I_{\rm R}$, we define 
\begin{equation*}
e_{\ell} \coloneqq \Theta^{\ast} - r_{\ell}, \ \ 
{\rm val}(\ell) \coloneqq \xi_{\ell}(a^{\ast}), \ \ 
{\rm cost}(\ell) \coloneqq \sum_{a \in A}\tau(a) \cdot \xi_{\ell}(a). 
\end{equation*}
For each element $\ell \in I_{\rm R}$, 
we define 
\begin{equation*}
N_{\ell} \coloneqq 
(\Theta^{\ast} + 1) \cdot {\rm val}(\ell) - {\rm cost}(\ell) - b(X_{\ell}). 
\end{equation*}
For each element $\ell \in I_{\rm R}$, 
we define the function ${\rm sg}_{\ell} \colon S \cup T \to \{-1,0,1\}$ by 
\begin{equation*}
{\rm sg}_{\ell}(v) \coloneqq 
\begin{cases}
1 & \mbox{if $v \in S \cap X_{\ell}$}\\
-1 & \mbox{if $v \in T \cap X_{\ell}$}\\
0 & \mbox{otherwise}. 
\end{cases}
\end{equation*}
Then for every integer $\ell \in I_{\rm R}$, 
\begin{equation*}
\begin{split}
N_{\ell} & = (\Theta^{\ast} + 1) \cdot {\rm val}(\ell) - {\rm cost}(\ell) - b(X_{\ell})\\
& = \sum_{a \in A}\varepsilon_{\xi_{\ell}}(a^{\ast},a) \cdot ((\Theta^{\ast}+ 1) \cdot c(a))
- \sum_{(a,f) \in A^2}\varepsilon_{\xi_{\ell}}(a,f) \cdot (\tau(a)\cdot c(f))
- \sum_{v \in S \cup T}{\rm sg}_{\ell}(v) \cdot |b(v)|.  
\end{split} 
\end{equation*}
Notice that 
$(\Theta^{\ast} + 1) \cdot c(a) \ge 0$ for every arc $a \in A$, 
$\tau(a)\cdot c(f) \ge 0$ for every pair $(a,f) \in A^2$, and 
$|b(v)| \ge 0$ for every vertex $v \in S \cup T$. 
Furthermore, these terms do not depend on $\ell$. 

\begin{lemma} \label{lemma:positive}
For each element $\ell \in I_{\rm R} \setminus \{\Pi\}$, 
we have ${\rm val}(\ell) > 0$ and 
\begin{equation*}
0 < e_{\ell} \cdot {\rm val}(\ell) \le 
N_{\ell} < 2 \cdot e_{\ell} \cdot {\rm val}(\ell). 
\end{equation*}
\end{lemma}
\begin{proof}
Let $\ell$ be an element in $I_{\rm R} \setminus \{\Pi\}$.
Since $\ell \neq \Pi$, 
$e_{\ell} > 0$.

Since $d(\theta_{\ell}) < 0$, 
$\Phi_{\theta_{\ell}}(X_{\ell}) < 0$. 
Thus, 
since $o_{\theta_{\ell}}(X_{\ell}) \ge 0$, 
$b(X_{\ell}) > 0$. 
This implies that 
\begin{equation} \label{eq_1:lemma:positive}
0 < b(X_{\ell}) \le o_{r_{\ell}}(X_{\ell}) = \rho_{X_{\ell}}(r_{\ell}) =  
\langle k_{r_{\ell}}^{X_{\ell}}, \xi_{\ell} \rangle
= 
(r_{\ell} + 1) \cdot {\rm val}(\ell) - {\rm cost}(\ell). 
\end{equation}
Since ${\rm cost}(\ell) \ge 0$, 
\eqref{eq_1:lemma:positive} implies that 
${\rm val}(\ell) > 0$. 
Furthermore, \eqref{eq_1:lemma:positive} implies that 
\begin{equation*}
\begin{split}
N_{\ell} 
& = (\Theta^{\ast} + 1) \cdot {\rm val}(\ell) - {\rm cost}(\ell) - b(X_{\ell})\\
& \ge (\Theta^{\ast} + 1) \cdot {\rm val}(\ell) - {\rm cost}(\ell) 
- (r_{\ell} + 1 ) \cdot {\rm val}(\ell) + {\rm cost}(\ell) \\
& = (\Theta^{\ast} - r_{\ell}) \cdot {\rm val}(\ell) 
= e_{\ell} \cdot {\rm val}(\ell) > 0.
\end{split}
\end{equation*}

What remains is to prove  
$N_{\ell} < 2 \cdot e_{\ell} \cdot {\rm val}(\ell)$. 
Since $r_{\ell}$ is the optimal solution to $\mbox{\sc qf}(X_{\ell})$, 
\begin{equation*}
b(X_{\ell}) > o_{r_{\ell}-1}(X_{\ell}) \ge 
\langle k_{r_{\ell}- 1}^{X_{\ell}}, \xi_{\ell} \rangle
= 
r_{\ell} \cdot {\rm val}(\ell) - {\rm cost}(\ell). 
\end{equation*}
Thus, since $e_{\ell}$ is a positive integer, 
\begin{equation*}
\begin{split}
N_{\ell} 
& = (\Theta^{\ast} + 1) \cdot {\rm val}(\ell) - {\rm cost}(\ell) - b(X_{\ell})\\
& < (\Theta^{\ast} + 1) \cdot {\rm val}(\ell) - {\rm cost}(\ell) 
- r_{\ell} \cdot {\rm val}(\ell) + {\rm cost}(\ell) \\
& 
= (\Theta^{\ast} - r_{\ell} + 1) \cdot {\rm val}(\ell) 
\le 2(\Theta^{\ast} - r_{\ell}) \cdot {\rm val}(\ell) 
= 2 \cdot e_{\ell} \cdot {\rm val}(\ell).
\end{split}
\end{equation*}
This completes the proof. 
\end{proof}

\begin{lemma} \label{lemma:reject}
$|I_{\rm R}| \le 3(p+1) \cdot \kappa(m) \cdot \kappa(M) + 1$, where 
we define $M \coloneqq m^2 + m + k$. 
\end{lemma}
\begin{proof}
Recall that, for each integer $\ell \in I_{\rm R}$, 
\begin{equation*}
{\rm val}(\ell) = \xi_{\ell}(a^{\ast}) = \sum_{a \in A}\varepsilon_{\xi_{\ell}}(a^{\ast},a) \cdot c(a).
\end{equation*}
Thus, since 
Lemma~\ref{lemma:positive} implies that 
\begin{equation*}
\{{\rm val}(\ell) \mid \ell \in I_{\rm R} \setminus \{\Pi\}\} 
\subseteq 
\{\langle z, c \rangle \mid \langle z, c \rangle > 0, z \in \{-1,0,1\}^A\}, 
\end{equation*}
there exist 
positive reals $h_1,h_2,\dots,h_{\delta}$ 
such that $\delta \le \kappa(m)$ and 
\begin{equation*}
\{{\rm val}(\ell) \mid \ell \in I_{\rm R} \setminus \{\Pi\}\} 
\subseteq \bigcup_{i=1}^{\delta}\{x \in \mathbb{R}_+ \mid h_i/2 < x \le h_i\}.
\end{equation*} 

Let $\gamma$ be an integer in $[\delta]$. 
Then we prove that 
\begin{equation} \label{eq_1:lemma:reject} 
|\{\ell \in I_{\rm R} \setminus \{\Pi\} \mid h_{\gamma}/2 < {\rm val}(\ell) \le h_{\gamma}\}|
\le 3(p+1) \cdot \kappa(M). 
\end{equation}
If we can prove \eqref{eq_1:lemma:reject}, then 
\begin{equation*}
|I_{\rm R} \setminus \{\Pi\}| 
\le 
\sum_{i=1}^{\delta}
|\{\ell \in I_{\rm R} \setminus \{\Pi\} \mid h_i/2 < {\rm val}(\ell) \le h_i\}|\\
\le 
3(p+1) \cdot \kappa(m) \cdot \kappa(M). 
\end{equation*}
Since $\Pi \in I_{\rm R}$, this implies the claimed bound. 

What remains is to prove \eqref{eq_1:lemma:reject}. 
If the set in the left-hand side of \eqref{eq_1:lemma:reject}
is empty, then \eqref{eq_1:lemma:reject}
clearly holds. Thus, we assume that the set in the left-hand side of \eqref{eq_1:lemma:reject} is not empty and 
\begin{equation*}
\{\ell \in I_{\rm R} \setminus \{\Pi\} \mid h_{\gamma}/2 < {\rm val}(\ell) \le h_{\gamma}\}
= \{x_1,x_2,\dots,x_\sigma\}, \ \ 
x_1 < x_2 < \dots < x_{\sigma}.  
\end{equation*}

\begin{claim} \label{claim_1:lemma:reject}
For every integer $j \in [\sigma - 1]$, 
\begin{equation*}
e_{x_{j+1}} \le \frac{p}{p+1} e_{x_j}.
\end{equation*}
\end{claim}
\begin{proof}
For every integer $j \in [\sigma - 1]$, 
since $\theta_{x_{j+1}} \ge r_{x_j}$, Lemma~\ref{lemma:geometric} implies that 
\begin{equation*}
e_{x_{j+1}} = 
\Theta^{\ast} - r_{x_{j+1}} 
\le 
\frac{p}{p+1}(\Theta^{\ast} - \theta_{x_{j+1}})
\le 
\frac{p}{p+1}(\Theta^{\ast} - r_{x_j})
=
\frac{p}{p+1}e_{x_j}.
\end{equation*}
This completes the proof. 
\end{proof}

Define $\omega \coloneqq 3(p+1)$. 
Then for each integer $i \in [\omega-1]_0$, we define 
\begin{equation*}
J_i \coloneqq \{x_j \mid j \in [\sigma], \ j \bmod \omega = i\}, \ \ 
\widehat{J}_i \coloneqq \{N_j \mid j \in J_i\}.
\end{equation*}
In what follows, we prove that 
$|J_i| \le \kappa(M)$ for every integer $i \in [\omega-1]_0$, 
which completes the proof of \eqref{eq_1:lemma:reject}. 

Let $i$ be an integer in $[\omega-1]_0$,
and let $x_j,x_{j+\omega}$ be elements in $J_i$. 
Claim~\ref{claim_1:lemma:reject} implies that 
\begin{equation*}
e_{x_{j+\omega}} 
\le \Big(\frac{p}{p+1}\Big)^{j+\omega - j}e_{x_j}
\le \Big(\frac{p}{p+1}\Big)^{3(p+1)}e_{x_j}
\le \frac{e_{x_j}}{8},
\end{equation*}
where the last inequality follows from the following well-known inequality.
\begin{equation*}
\Big(1 + \frac{1}{p}\Big)^{p+1} 
\ge 
\Big(1 + \frac{p+1}{p}\Big)
\ge 2.  
\end{equation*} 
Furthermore, 
since $x_j,x_{j+\omega} \in J_i$, ${\rm val}(x_{j+\omega}) \le 2 \cdot {\rm val}(x_j)$.
Thus, Lemma~\ref{lemma:positive} implies that 
\begin{equation*}
2N_{x_{j+\omega}} 
< 2 \cdot 2 \cdot e_{x_{j+\omega}} \cdot {\rm val}({x_{j+\omega}})
\le 4 \cdot \frac{e_{x_j}}{8} \cdot 2 \cdot {\rm val}({x_j})
\le e_{x_j} \cdot {\rm val}({x_j})
\le N_{x_j}. 
\end{equation*}
Thus, since this implies that 
$N_{x_j} \neq N_{x_{j+\omega}}$, 
we have $|J_i| = |\widehat{J}_i|$.
Furthermore, Lemma~\ref{lemma:Radzik13}
implies that $|\widehat{J}_i| \le \kappa(M)$. 
This completes the proof. 
\end{proof} 

Lemmas~\ref{lemma:accept} and \ref{lemma:reject} imply 
that Algorithm~\ref{alg:main} is a polynomial-time 
algorithm. 
In what follows, we derive an explicit running-time bound
for Algorithm~\ref{alg:main}. 
Lemmas~\ref{lemma:accept} and \ref{lemma:reject} imply 
that the number of iterations in Algorithm~\ref{alg:main}
is $\widetilde{O}(k^2m^3)$, 
where $\widetilde{O}(\cdot)$ denotes the Landau symbol
omitting all poly-logarithmic factors. 
Step~3 can be done 
in $\widetilde{O}(k^3m^2)$ time 
by using the submodular function minimization algorithm by \cite{LeeSW15} and 
the minimum-cost flow algorithm of \cite{Orlin93} 
as an oracle. 
Step~4 can be done 
in $\widetilde{O}(m^2n)$ time~\cite{SahoS17}. 
Finally, the total time complexity is 
$\widetilde{O}(k^5m^5 + k^2m^5n)$.
It should be noted that this time complexity is worse than 
$\widetilde{O}(k^5m^2 + k^3m^3 + m^3n)$ in \cite{SchloterST22}. 

\bibliographystyle{plain}
\bibliography{simple_discrete_newton_bib.bib}

\end{document}